# Application of HUBzero platform for the educational process in astroparticle physics

**Yulia Kazarina[1,a], Igor Bychkov[1,2], Alexander Kryukov[3], Julia Dubenskaya[3], Elena Korosteleva[3], Minh-Duc Nguyen[3], Stanislav Polyakov[3], Evgeny Postnikov[3], Andrey Mikhailov[2], Alexey Shigarov[2], Oleg Fedorov[1], Dmitry Shipilov[1], Dmitry Zhurov[1]**

*[1] Applied Physics Institute, Irkutsk State University, Irkutsk, Russia*

*[2] Matrosov Institute for System Dynamics and Control Theory, Siberian Branch of Russian Academy of Sciences, Irkutsk, Russia*

*[3] Lomonosov Moscow State University, Skobeltsyn Institute of Nuclear Physics, Moscow, Russia*

E-mail: [a] kazarina@astroparticle.online

In the frame of the Karlsruhe-Russian Astroparticle Data Life Cycle Initiative it was proposed to deploy an educational resource astroparticle.online for the training of students in the field of astroparticle physics. This resource is based on HUBzero, which is an open-source software platform for building powerful websites, which supports scientific discovery, learning, and collaboration. HUBzero has been deployed on the servers of Matrosov Institute for System Dynamics and Control Theory. The educational resource astroparticle.online is being filled with the information covering cosmic messengers, astroparticle physics experiments and educational courses and schools on astroparticle physics. Furthermore, the educational resource astroparticle.online can be used for online collaboration. We present the current status of this project and our first experience of application of this service as a collaboration framework.

# 1. Introduction

The rapid development of the astroparticle physics for more than a hundred years of its existence was associated both with the rapid development of various areas of classical, quantum and relativistic physics, on the one hand, and with the construction of large telescopes, the emergence of fundamentally new radiation detectors and computerized methods for processing observations, on the other. All this makes astrophysical studies invaluable for solving fundamental physical problems. Modern scientific research in the field of astroparticle physics is inextricably linked with the conduct of complex calculations and the use of high-performance computing resources. At the same time, the effectiveness of research directly depends on the availability and accessibility of computing applications for solving one or another problem. Currently, a large baggage of such applications has been accumulated, including libraries of numerical methods, applied computational packages, computational models, etc. The researcher needs to independently install and configure the application on his computer or computational resource. In the case of a new version of the application, these actions must be repeated again. In addition, in the case of parallel applications, the researcher also needs to master the mechanism for launching the application on a computational resource accessible to it. Often there is a situation where the researcher can correctly formulate the task, but does not have the necessary qualifications to perform the described actions. To solve these problems, high-level environments are used that provide researchers with remote access to computing applications through problem-oriented interfaces. As a rule, these environments are implemented in the form of web portals, and work with applications is carried out through a web browser. The availability of ready-made software will allow young researchers in most cases to avoid the time-consuming implementation of the program code and concentrate on the problem being solved, and qualified researchers will give the opportunity to give colleagues access to their developments. More than that such as a web portal can be filled by relevant information necessary for scientific research.

## 1.1. Objects and goals

The objects of research and development in this paper are the methods of presenting scientific knowledge in the field of astroparticle physics using Internet technologies. The purpose of this work is to deploy the HUBZero platform [1-4] for filling it with educational materials for the preparation of students in the field of astroparticle physics, such as educational courses, documentation and exercises on MC simulation, examples of data analysis, including collaboration tools like simulation and data analysis software, astroparticle forum. Such an educational resource has been deployed on the servers of Matrosov Institute for System Dynamics and Control Theory on the basis of the open-source HUBzero platform and named astroparticle.online [5].

## 1.2. Content

Firstly, it is proposed to provide on astroparticle.online the actual information about the theoretical aspects of astroparticle physics, including the main astroparticle messengers: cosmic rays, gamma-astronomy, neutrino astronomy and gravitational waves. Also, the description of the modern Russian astroparticle physics experiments, TAIGA observatory [6] and GVD neutrino experiment [7], is presented there. Besides, the educational resource astroparticle.online contains the information about current schools in astroparticle physics, as well as a calendar of the most important international conferences and seminars in this field of research.

## 1.3. First experience

The first experience of using the educational resource astroparticle.online was obtained at the the ISAPP-Baikal Summer School "Exploring the Universe through multiple messengers" [8], created jointly by Irkutsk State University and the Joint Institute for Nuclear Research (Dubna) as part of a cooperation agreement. The main idea of the school is the education and training of students and young scientists in the modern aspects of elementary particle physics and astrophysics by leading physicists, theorists and experimenters, as well as the creation of a contact base and the choice of the direction of research by young specialists. The school is regularly held in the small village of Bolshie Koty on the shore of Lake Baikal. Due to lack of internet connection in the village it was proposed to deploy the astroparticle.online locally. This allowed the organizers of the school to spread the conference materials, lectures and student reports online. Also, the participants could post their impressions (photos and videos), comments on the page of the school. The screen of the school page is presented at the Figure 1.

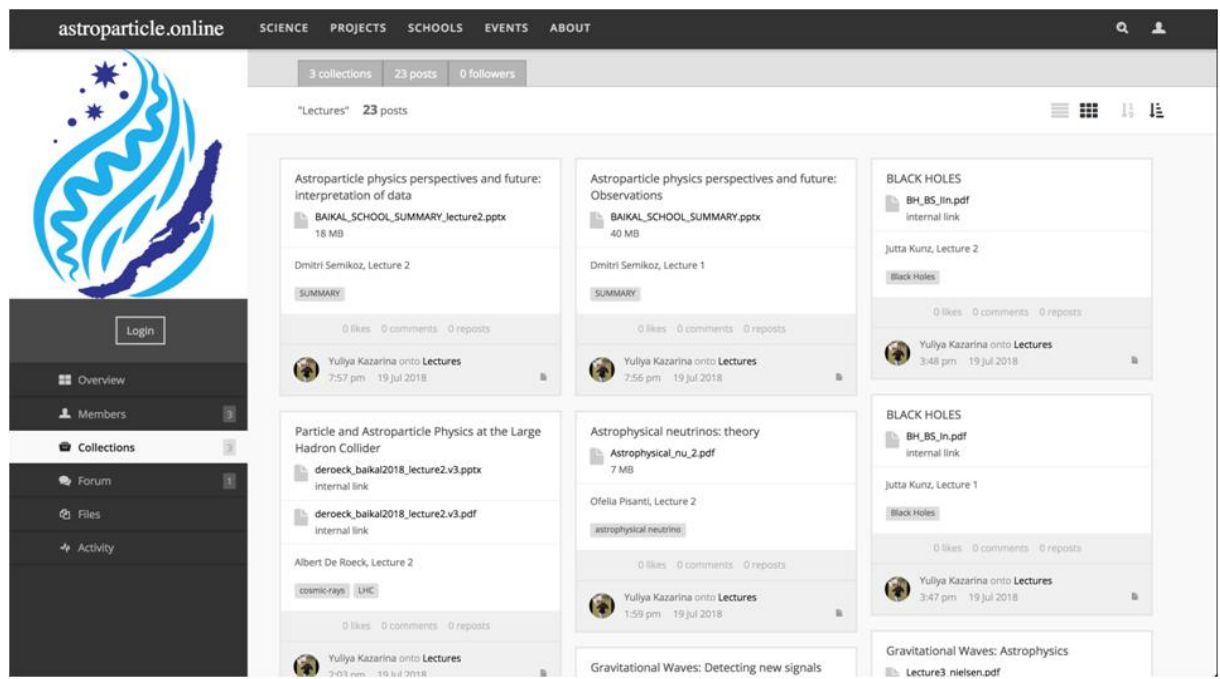

Figure 1. Screen of the school page on educational resource astroparticle.online.

## 2. Technical aspects

HUBzero - is an open-source software platform for building powerful Websites that support scientific discovery, learning, and collaboration. The HUBzero software allows individuals to access simulation tools and share information. HUBzero has several advantages: everything is customized for the needs of the collaboration and it is possible to expand the functionality.

HUBzero built from open-source software: the Linux operating system (Debian or RHL), the Apache web server, the MySQL database, the Joomla content management system, and the PHP web scripting language, exim4 mail server, subversion system.

**2.1. Installation**

There are three ways to get instance of HUBzero: manual install all modules and packages, install from ready virtual image (vmdk) or install from ready build for AWS. Minimum system requirements for each type of installation: Linux Debian 7 or 8, 4gb of RAM, 2 core processor, 20gb HDD.

The second type of the installation has been chosen, because it allows one to deploy the system quickly if necessary. Also this eliminates the need for manual reconciliation of different package versions. This installation requires a pre-configured virtual machine (VMware instance). In our case we used VMware vSphere. This server-side application allows one to create virtual machines quickly, allocate them the necessary size of memory and CPU recourse. When creating a virtual machine, it should be explicitly stated that the vmdk image will be used as the source for the system. During the installation problems with the type of image could be occur, as newer versions of vSphere use "thick" type of vdmk, but Hubzero official site provides only "thin" format. Therefore, the "thin" to "thick" conversion should be used using the vSphere console tool-kit.

After installation, the network interfaces should be configured - set a static address, gateway address and network mask address. These parameters are set in the file

/etc/network/interface.

In some cases, it is required to register separate routes if there are any features associated with an external firewall, or a gateway, or other features of the network topology (NAT, etc.) It is also necessary to change the passwords for root user, database and CMS for security reasons. Next, the settings for pages, modules, user rights and directories used by the site can be configured.

**2.2. Cloning the instance and possible bugs**

To use the HUBzero system at the ISAPP-Baikal Summer School, it was necessary to solve some of problems. First of all, there is no stable Internet connection on Lake Baikal, therefore it is necessary to deploy an up-to-date copy of the site accessible via a local wireless network. Then, it is important to solve the inverse problem - to make all the information that was posted within the school, accessible to all on the Internet.

It was decided to use the following method - creating and deploying a backup, directly CMS HUBzero (without cloning whole VM image), since the other components of the system remain unchanged. Everything in the CMS is stored in two entities: a database where information about users, groups, rights, passwords, text and graphic information, and files of the system itself, which are responsible for the appearance of the site, user interaction interfaces, and so on, is stored.

To transfer, the full backup of the database and CMS files should be made. A backup copy of the database is created using the command line, using the console program mysqldump, which allows one to create full dumps of the required tables, when started with the following syntax

"mysqldump –u user –p password base name> dump.sql".

Further on the target machine, this dump can be deployed using the mysql program.

To save the same CMS file, the compression using the tar program was used, which allows one to create archives, as well as use compression. An important feature here is that with such a compression, all rights to the directories that were configured earlier are retained. Next, the image is deployed on the target machine, also using the tar utility.

In rare cases, errors are possible if some of the files were not completely saved, or the archive was damaged when downloading or placing files, to check the checksum using the md5 program is necessary to ensure that the integrity of the data is fully observed. Errors are also possible when deploying a database from a backup. The mysql program indicates the existence of certain tables, although they have not yet been created.

# 3. Future plans

In the near future it is planned that there will be created the online educational course on the experimental astroparticle physics in the frame of the educational resource astroparticle.online. The course will be supplemented with examples of real and simulated astrophysical data, will contain tools for processing and analyzing these data, as well as documentation and exercises for Monte Carlo simulation. For the students who are involved in the TAIGA and GVD activity there will be provided the astroparticle forum as the first step of the collaboration framework.

# 4. Conclusion

The educational resource astroparticle.online has been deployed on the servers of Matrosov Institute for System Dynamics and Control Theory on the basis of the open-source HUBzero platform. The project is positioned as an online educational platform for students in the direction of astroparticle physics. It is being filled with educational materials for preparing students in the field of astroparticle physics, as well as improving the quality of educational material. The resource was successfully tested at the ISAPP-Baikal Summer school. The technical aspects of the installation, configuration and cloning are also presented in the paper.

## Acknowledgments

This work was financially supported by Russian Science Foundation and Helmholtz Society, Grant No. 18-41-06003. The developed educational resources were freely deployed on the cloud infrastructure of the Shared Equipment Center of Integrated Information and Computing Network for Irkutsk Research and Educational Complex (http://net.icc.ru).